\documentclass{aa}  
\usepackage{natbib}
\bibpunct{(}{)}{;}{a}{}{,} 
\usepackage{graphicx}
\usepackage{txfonts}
\begin{document} 

   \title{The dark matter distribution in the spiral $NGC~3198$ out to 0.22 $R_{vir}$}


   \author{E.V. Karukes\inst{1,2},  P. Salucci\inst{1,2}
          \and
          G. Gentile \inst{3,4}
         }     
   \institute{SISSA/ISAS, International School for Advanced Studies, Via
  Bonomea 265, 34136, Trieste, Italy \\
              \email{ekarukes@sissa.it}
              \and
INFN, Sezione di Trieste, Via Valerio 2, 34127, Trieste, Italy       
                 \and
 Department of Physics and Astrophysics, Vrije Universiteit Brussel, Pleinlaan 2, 1050 Brussels, Belgium                
        \and
Sterrenkundig Observatorium, Universiteit Gent, Krijgslaan 281, 
  B-9000 Gent, Belgium
             }
      
\authorrunning{E.V. Karukes et al.}

 
 
\abstract
 {}  
  {We use recent very extended HI kinematics (out to 48 kpc) along with previous H$\alpha$ kinematics of the spiral galaxy NGC 3198 in order to derive its distribution of dark matter (DM).}  
   {First, we used a chi-square method to model the rotation curve (RC) of this galaxy in terms of different profiles of its DM distribution: the universal rotation curve (URC) mass model (stellar disk $+$ Burkert halo $+$ gaseous disk), the NFW mass model (stellar disk $+$ NFW halo $+$ gaseous disk) and the baryon $\Lambda$CDM mass model (stellar disk $+$ NFW halo modified by baryonic physics $+$ gaseous disk). Second, to derive the DM halo density distribution, we applied a new method that does not require a global and often uncertain  mass modelling.}
   {While according to the standard method, both URC and NFW mass models can account for the RC, the new method instead leads to a density profile that is sharply disagrees with the dark halo density distribution predicted within  the Lambda cold dark matter ($\Lambda$CDM) scenario.  We  find that  the effects of baryonic physics  modify the original  $\Lambda$CDM halo densities in such a way that  the resulting profile is more compatible with the DM density of NGC 3198 derived using our new method. However, at large distances, r $\sim$ 25 kpc, also this modified baryon $\Lambda$CDM halo profile appears to create a tension with the derived DM halo density.}
 {}
 
 \maketitle
%

\section{Introduction}
It has been known for several decades that the kinematics of disk galaxies leads to a mass discrepancy \citep[e.g.][]{bosma78, bosma79, rubin}. While in their inner regions that range between one and three disk exponential scale lengths according to the galaxy luminosity \citep{PS99},  the observed baryonic matter accounts for the rotation curves (RCs) \citep[e.g.][]{athanassoula, perssal, palunas}, we must add an extra mass component in the outer regions, namely a dark matter (DM) halo to account for that component. The kinematics of spirals is now routinely  interpreted in the framework of a DM component. In the widely accepted Lambda cold dark matter ($\Lambda$CDM) scenario, the virialized structures are distributed according the well known NFW profile proposed by Navarro, Frenk, and White \citep{nfw}. The $\Lambda$CDM scenario describes the large-scale structure of the Universe well \citep[e.g.][]{springel}, but it seems to fail on the scales of galaxies  \citep{blokbosma, gentile04, gentile05}. Going into detail, the NFW density profile leads to the ``core-cusp problem'':  empirical profiles with a central core of constant density, such as the pseudo-isothermal  \citep{iso1,iso2}  and the Burkert  \citep{salucciburkert2000}, fit the available RCs  much better than the mass models based on NFW haloes.
\newline
\indent
In the present paper, we derive the DM content and distribution in the spiral galaxy NGC 3198. This galaxy has been the subject of several investigations. It was studied by means of optical  \citep{cheriguene, hunter, bottema, waverskruit, kent, corradi, daigle} and HI-21 cm radio observations \citep[][the latter established it as the object with the clearest evidence for DM]{bosma81,vanalbada,begeman}, see also \citep{bloketal, gentile}. 

Our present analysis is mainly based on the HI observations  by \citet{gentileetal}, part of the HALOGAS (Westerbork Hydrogen Accretion in LOcal GAlaxieS) survey. The main goal of HALOGAS is to investigate the amount and properties of extra-planar gas by using very deep HI observations. In fact, for this galaxy, they present a very extended RC out to 720 arcsec, corresponding to $\sim$ 48 kpc for a galaxy distance  at 13.8 Mpc \citep{freedman}. The previous HI observations by \citet{bloketal} were only extended out to $\sim$ 38 kpc, for the same galaxy distance. In \citet{gentileetal} this extended RC  was modelled  in the framework of modified Newtonian dynamics (MOND). Here, we want to use such a uniquely extended kinematics to help resolving  the  DM core-cusp issue. To the very reliable  kinematics available from 2 to 48  kpc we apply {\it  two} different mass decomposition  methods  that will derive the DM halo structure. This is compared with a) the empirically based halo profiles coming from the URC, b)  the NFW haloes and c) the baryon $\Lambda$CDM haloes, the outcome of scenarios in which baryonic physics has shaped the DM halo density.

\begin{figure*}
\begin{centering}
\includegraphics[angle=0,height=8truecm,width=13truecm]{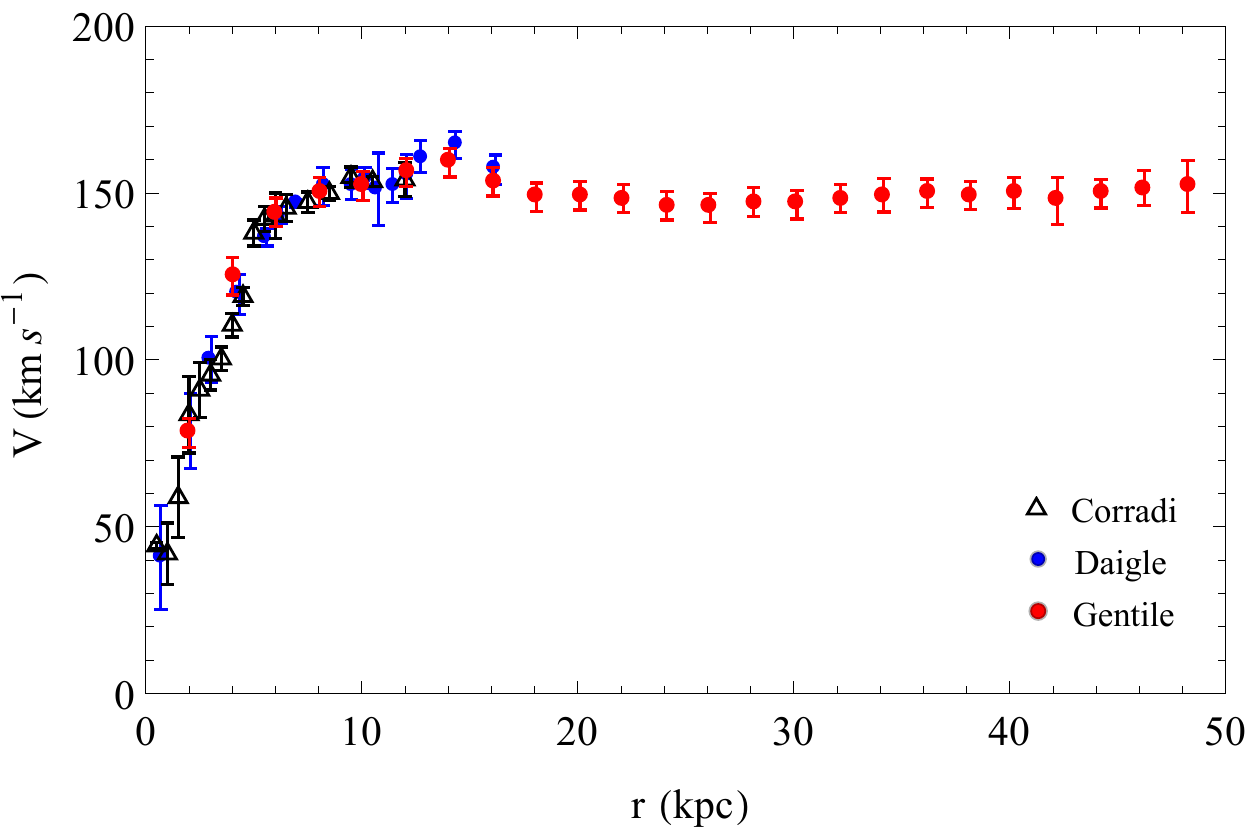}
\caption{Comparison between H$\alpha$ and HI RCs black open triangles with error bars from \citet{corradi}, blue circles with error bars are from  \citet{daigle}, and red circles with error bars are from \citet{gentileetal}.}
\label{DATA}	 
\end{centering}
\end{figure*} 

This paper is organized as follows. In Sect. 2 we present the HI and H$\alpha$ kinematics used in this study. In Sect. 3 we model the RC by using the quadrature sum of the contributions of the individual mass components (stellar disk $+$ dark halo $+$ gas disk) where the  dark halo has a NFW or a Burkert density profile, respectively. In Sect. 4 we obtain  the results of standard  mass modelling of the NGC 3198 RC. In Sect. 5 we use a recent mass modelling  technique to obtain a very robust and careful determination of the DM halo density of NGC 3198 and to show that it is at variance with the NFW density profile in an unprecedented way. We consider the  mass dependent density profiles obtained by hydro-dynamical simulations of $\Lambda$CDM  haloes in Sect. 6. Our conclusions are drawn in Sect. 7. 

    \begin{table}
\centering
 \caption{Stellar disk contribution  $V_{d}$  (km $\mathrm{s^{-1}}$) \citep{bloketal} and the circular velocity $V$  (km $\mathrm{s^{-1}}$) mainly from \citet{gentileetal}, but also from  \citet{corradi} and \citet{daigle}, of the NGC 3198 with errors d$V$ (km $\mathrm{s^{-1}}$). The DM density profile $\rho\times{10^{-25}}$ (g $\mathrm{cm^{-3}}$) (see Sect. 5 for details).}
 
 \label{tbl:1}
  \begin{tabular}{  c c c c c}  
    \hline
     \hline\\   
	    R  &  $V_{d}$ & $V$&   d$V$  & $\rho\times{10^{-25}}$ \\ (kpc)  & (km $\mathrm{s^{-1}}$)   &  (km $\mathrm{s^{-1}}$)  &  (km $\mathrm{s^{-1}}$) &  (g $\mathrm{cm^{-3}}$) \\ \hline\\
    
      2.0 &   86.2   &   79.0 &  7.0& ---\\   
      3.0 &   85.4  &   97.8  &   5.0&---\\
     4.0 & 93.6   & 118.0 & 5.6&--- \\ 
      5.5 &  115.7 & 139.4 &  4.3&2.34\\
     6.0 & 120.8  & 144.2 & 4.3&2.33 \\ 
     7.0 &  125.4 &  143.3 & 4.5&2.20\\
    8.0 & 125.5  & 150.3 & 4.3&2.01 \\ 
     9.0 &  123.5 &  149.9  & 4.3&1.83 \\
  10.1 & 120.1 &  152.1 & 4.3&1.64 \\ 
   11.0 & 116.6 &  151.1  & 4.5&1.48\\
  12.1 & 112.6 &  156.2 & 4.3&1.32 \\ 
   14.1 & 105.2 &  161.0 & 4.3&1.06 \\ 
   16.1 & 98.6  &  155.3 & 4.3&0.86\\ 
  18.1 & 92.7  &  148.7& 4.3&0.70 \\ 
  20.1 & 87.5  &  149.1 & 4.3&0.58 \\ 
  22.1 & 82.8  &  148.4 & 4.3&0.48\\
  24.1 & 78.7  &  146.2 & 4.3&0.42 \\
  26.1 & 75.1  &  145.5 &  4.3&0.36 \\
  28.1 & 71.9  &  147.3 & 4.3&0.33 \\
  30.2 & 68.9  &  146.5 & 4.3&0.30 \\
  32.2 & 66.3  &  148.4 & 4.3&0.27  \\
  34.2 & 63.9  &  149.3 & 5.0&0.25 \\
  36.2 & 61.8  &  149.9 & 4.3&0.23 \\
  38.2 & 59.8  & 149.3 & 4.3&0.21  \\
  40.2 & 58.0  &  150.0 & 4.6 &0.20\\
  42.1 & 56.4  &  147.6 & 7.0 &0.18\\
  44.2 & 54.9  &  149.8 & 4.3 &0.16\\
  46.2 & 53.5  &  151.5 & 4.3&0.13  \\
  48.2 & 52.2  &  151.9 & 7.7 &0.11 \\  
            \hline
  \end{tabular}  
\end{table} 


\section{KINEMATICS DATA}

\subsection{HI data}

The HI data of NGC 3198 were taken in the framework of the HALOGAS survey \citep{heald} and they were presented in  \citet{gentileetal}. The
data were obtained with the WSRT (Westerbork Synthesis 
Radio Telescope) for 10 $\times$ 12 hours, with a total bandwidth of 10
MHz subdivided into 1024 channels. The data cube  used to derive the
rotation curve has a beam size of 35.2 $\times$ 33.5 arcsec, and we were
able to detect emission down to $\sim10^{19}$ atoms cm$^{-2}$. 

To construct the gas distribution and the rotation curve of this galaxy in a reliable way ,
we modelled the whole data cube by means of the TiRiFiC software (J\'ozsa et 
al. 2007). We successfully modelled the HI observations of NGC 3198 with a
thin and a thick neutral hydrogen disk, and,  thanks to an increased
sensitivity,  we were able to trace the rotation curve out to a 
distance of  $\sim$ 48 kpc (for the galaxy distance of 13.8 Mpc),  i.e. to a larger radius than those reached in 
previous studies. More details about the data  reduction analysis and modelling can be found in  \citet{gentileetal}.

\subsection{H$\alpha$ data}

  The H${\alpha}$ rotation curves of NGC 3198 have been published by many authors \citep{cheriguene, hunter, bottema, waverskruit, kent, corradi, daigle}. We notice that \citet{corradi} and \citet{daigle} measurements are a good representation of these data.  In Fig. \ref{DATA}, these RCs  are plotted  along with the HI RC used in this work. H$\alpha$ data provide us measurements of the circular velocity at five different new  (inner) radii not mapped by the present HI RC and four more measurements at radii at which we can combine them with our  HI data.  The hybrid RC  (HI+H$\alpha$ data) is listed in Table \ref{tbl:1}. All together,  HI and H$\alpha$  RCs agree very well  within the observational errors where they coexist. With respect to the HI RC of \citet{gentile}, the present hybrid RC  covers the innermost and the outermost regions significantly better; however, H$\alpha$ RCs give us no new useful information for $r<2$ kpc, owing to its large observational uncertainties and because in  this  very inner region the kinematics is strongly influenced by non-axisymmetric motions  \citep[see][]{corradi}.

\section{MASS MODELS}

We model the spiral galaxy NGC 3198 as consisting of three ``luminous'' components, namely the bulge and the stellar and the gaseous disks, which are embedded in a dark halo. To study the properties of luminous and dark matter in this galaxy, we model the RC in terms of the contributions from the stellar disk, the bulge, the gaseous disk, and the dark matter halo:

\begin{equation}
V^2(r)=V_{\mathrm{d}}^2(r)+V_{\mathrm{b}}^2(r)+V_{\mathrm{g}}^2(r)+V_{\mathrm{DM}}^2(r).
\label{vel}
\end{equation}

\subsection{Luminous matter}
\label{3.1} 
   
We define $V_{\mathrm{d}}(r)$ as the contribution of the stellar disk to the circular velocity $V(r)$. The surface brightness profile of NGC 3198 has been analysed very well by \citet{bloketal}. We assume their one-component surface brightness profile to derive  $V_{\mathrm{d}}(r)$. As a reference value we take the stellar mass-to-light ratio from \citet{bloketal}: $\Upsilon_{*,\mathrm{de Blok}}^{3.6}=0.8$ (referred to the Spitzer IRAC 3.6 $\mu$m band, which is a good proxy for the emission of the stellar disc). We thus set $V_\mathrm{d}^{2}(r)=(V_\mathrm{d}^{\mathrm{de Blok}}(r))^{2}\frac{\Upsilon_{*,\mathrm{fit}}^{3.6}}{0.8}$. Then, we leave the {\it amplitude} of the disk contribution to the circular velocity (i.e. the disk mass) as a free parameter to be derived by fitting the RC. No results of the present paper will change by assuming any other models of the dostribution of the stellar disk  of NGC 3198 in  \citet{bloketal} or  in previous works.

The contribution from the bulge component is $V_{\mathrm{b}}(r)$. We follow \citet{bloketal} and  we consider their 1-component model which accounts for this inner stellar component. Hence, $V_{\mathrm{b}}(r)=0$ since the bulge is included in $V_\mathrm{d}^{\mathrm{de Blok}}(r)$.

 The helium corrected contribution of the gaseous disk derived from the HI surface density distribution $V_{\mathrm{g}}(r)$ taken from \citet{gentileetal}. We notice that, thanks to the accuracy of the HI measurements and to the excellent knowledge of the distance of  this galaxy, this component is derived here much better than in the majority of the spirals  studied to resolve the core-cusp controversy.

The "luminous"  component of the circular velocity of this galaxy is then well known, except for the value of the disk mass which contributes to putting NGC 3198 at the front line of DM research.  

\subsection{Dark matter}
\label{3.2}

We define $V_{\mathrm{DM}}(r)=\int\limits_0^{r} 4 \pi \rho_{\mathrm{DM}} \mathrm{R}^2$dR as  the contribution to $V(r)$ from the dark matter halo of the virial mass $M_{vir}=\frac{4}{3} \pi 100 \rho_{crit} R_{vir}^{3}$ that could have a variety of density profiles. Here $\rho_{\mathrm{DM}}$ represents a DM halo density profile.

\subsubsection{Burkert halo}
\label{3.2.1}

The URC of galaxies and  the kinematics of individual  spirals  \citep{salucci2007} points to dark haloes with a central constant-density core, in particular, to the Burkert halo profile \citep{burkert, salucciburkert2000}. The relative density distribution is given by

\begin{equation}
\rho_\mathrm{URC}(r)=\frac{\rho_0 r_\mathrm{c}^3}{(r+r_\mathrm{c})(r^2+r_\mathrm{c}^2)}
\label{urc}
\end{equation}

\noindent
where $\rho_\mathrm0$ (the central density) and $r_\mathrm{c}$ (the core radius) are the two free parameters. The present data cannot distinguish these URC haloes from other {\it  cored} profiles, for which $\lim _{r\to 0}  \rho(r)  =  const$.

\subsubsection{Navarro, Frenk and White (NFW) halo} 
\label{3.2.2}

In numerical simulations performed in the ($\Lambda$) CDM scenario of structure formation, \citet{nfw} found that virialized systems follow a universal DM halo profile.  This is written as

\begin{equation}
\rho_\mathrm{NFW}(r)=\frac{\rho_\mathrm s}{(\frac{r}{r_s})(1+\frac{r}{r_s})^2}
\label{nfw}
\end{equation}

\noindent
where $\rho_\mathrm s$ and $r_s$ are the characteristic density and the scale radius of the distribution, respectively. These two parameters can be expressed in terms of the virial mass $M_\mathrm{vir}$, the concentration parameter $c=\frac{R_\mathrm{vir}}{r_s}$, and the critical density of the Universe $\rho_\mathrm{crit}=9.3 \times 10^{-30} \mathrm{g\,cm^{-3}}$. By using Eq. (\ref{nfw}), we can write

\begin{eqnarray}
\rho_{s} &=& \frac{100}{3} \frac{c^3}{\log{(1+c)}-\frac{c}{1+c}} \ \rho_\mathrm{crit}\quad \mathrm{g\,cm^{-3}},\nonumber\\
r_{s} &=& \frac{1}{c}\left(\frac{3 \times M_\mathrm{vir}}{4 \pi 100 \rho_\mathrm{crit}}\right)^{1/3} \quad \mathrm{kpc},
\label{rho}
\end{eqnarray} 

where $c$ and $M_\mathrm{vir}$ are not independent. It is well known from simulations that a $c-M_\mathrm{vir}$ relationship emerges \citep{klypin,bullock, wechsler}:

\begin{eqnarray}
c &\simeq& 11.7 \left(\frac{M_\mathrm{vir}}{10^{11}M_{\odot}}\right)^{-0.075}.
\label{cdm}
\end{eqnarray} 

We are testing the density profile of haloes made by collision-less cold dark matter particles. Variations in this scenario are not considered in this work, except one in Sect. 6.

\section{RESULTS FROM THE $\chi ^{2}$ FITTING METHOD}

 \begin{figure*}
\begin{centering}
\includegraphics[angle=0,height=8truecm,width=13truecm]{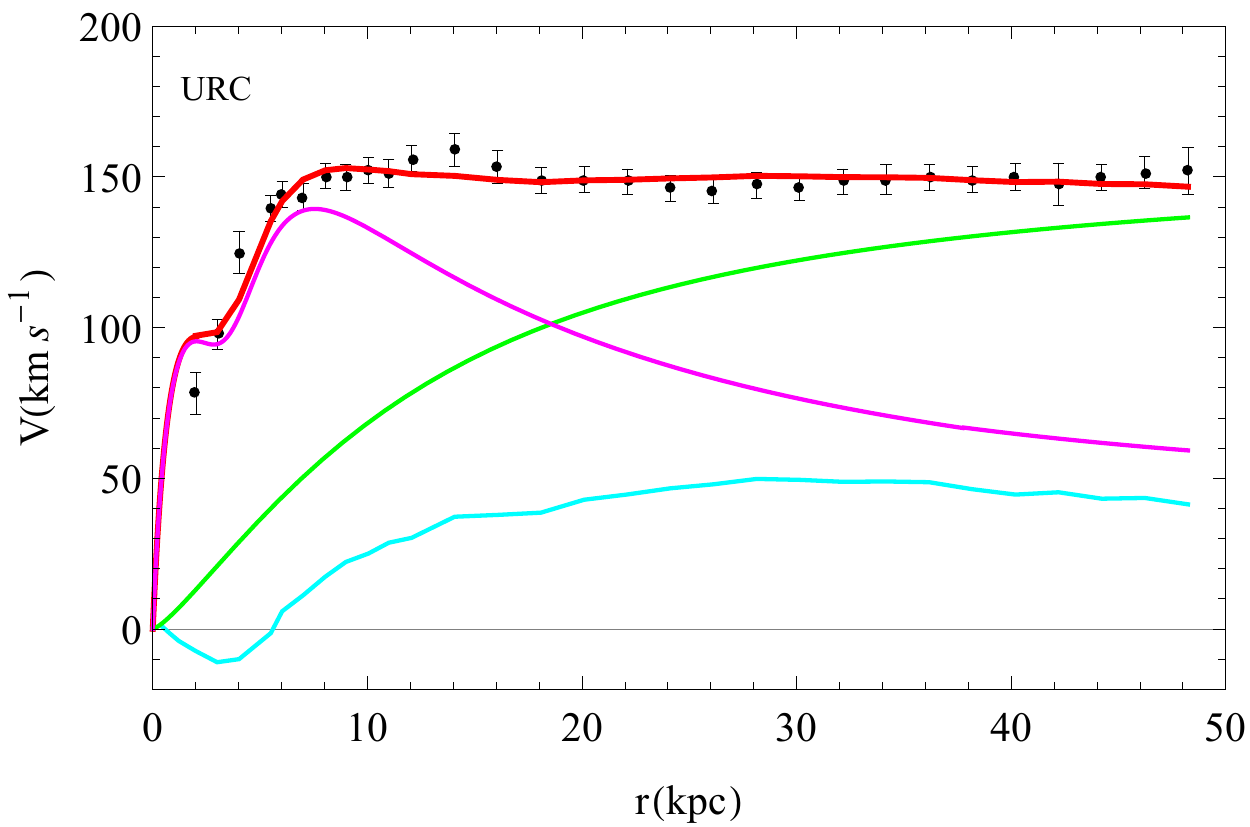}
\caption{URC mass modelling of NGC 3198. Circular velocity data
  (filled circles with error bars) are modelled (thick red line) by the halo cored component
  (thick green line), the  stellar disk (magenta line) and the HI disk (azure line).}
\label{BURKERT}	 
\end{centering}
\end{figure*} 

\begin{figure*}
\begin{centering}
\includegraphics[angle=0,height=8truecm,width=13truecm]{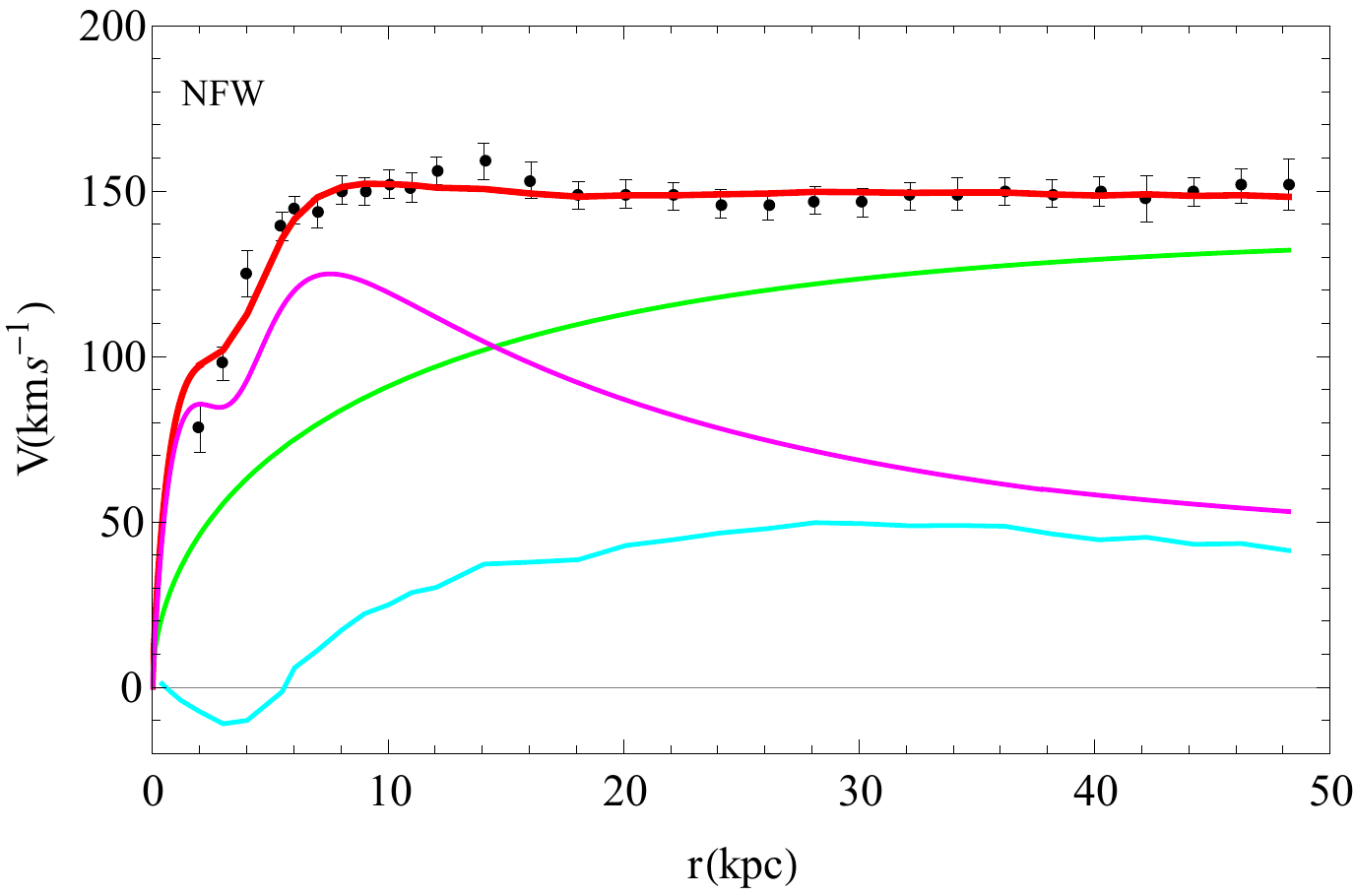}
\caption{NFW mass modelling of NGC 3198. Circular velocity data (filled circles with error bars) are modelled (thick red line) by the stellar disk (magenta line), the NFW halo profile (green line) and the HI contribution (azure line).}
\label{NFW}	 
\end{centering}
\end{figure*} 
 
The mass modelling results  for the Burkert and NFW profiles are shown in Figs (\ref{BURKERT}) and (\ref{NFW}), respectively.
The URC Burkert halo gives an excellent  fit (see  Fig. \ref{BURKERT}) with a reduced chi-square value of ${\chi}^2=0.9$. The best-fit parameters  are

\begin{eqnarray}
\rho_\mathrm{0} &=& (3.19 \pm 0.62) \times 10^{-25} \quad \mathrm{g\,cm^{-3}};\nonumber\\
r_\mathrm{c} &=& (17.7 \pm 2.0) \quad \mathrm{kpc};\nonumber\\
\Upsilon_*^{3.6} &=& (0.98 \pm 0.04).\nonumber
\end{eqnarray}

\begin{figure*}
\begin{centering}
\makebox[\textwidth][c]{
\includegraphics[width=0.208\textwidth, angle=0]{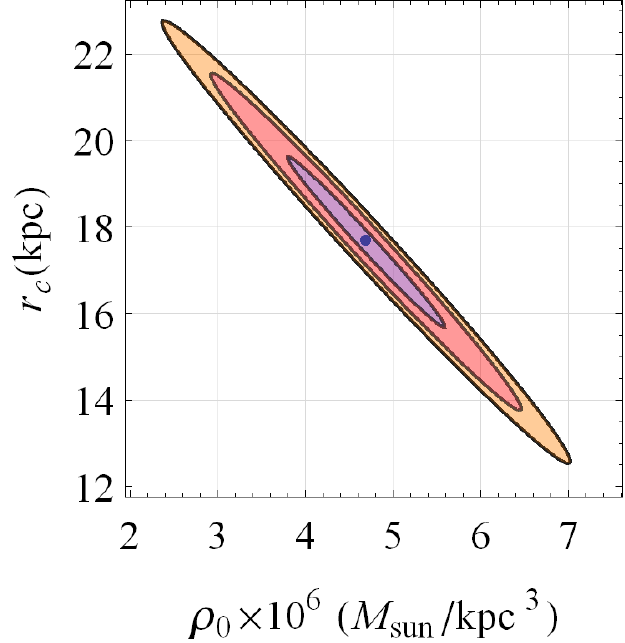}
\includegraphics[width=0.222\textwidth, angle=0]{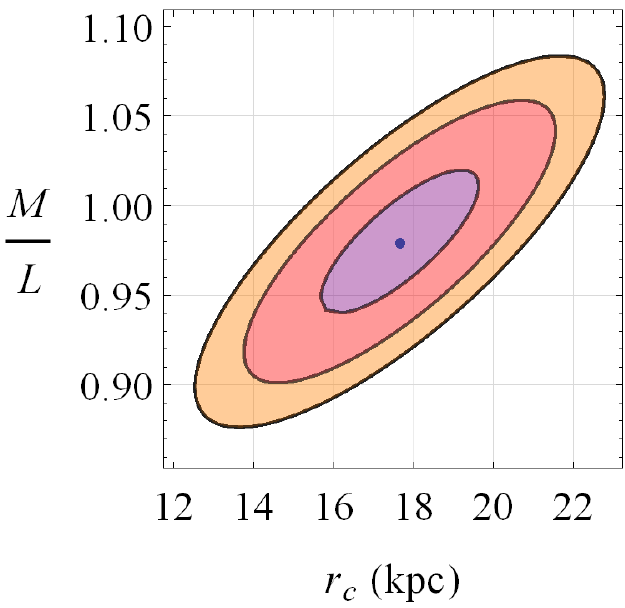}
\includegraphics[width=0.220\textwidth, angle=0]{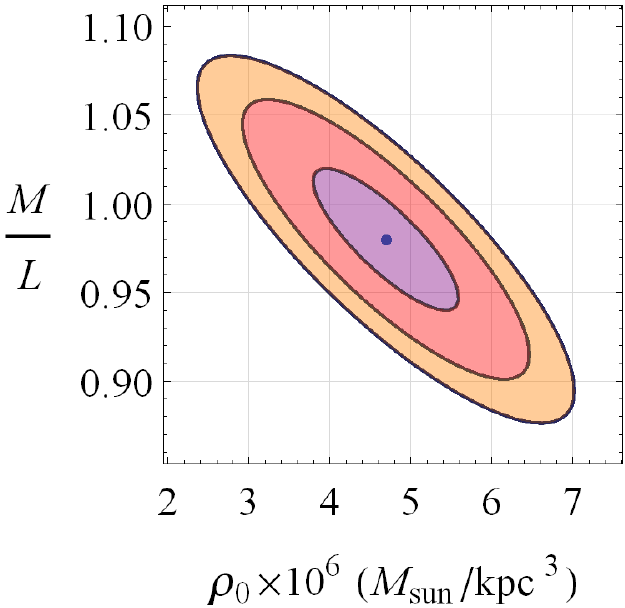}
}

\caption{1, 2, 3 $\sigma$ confidence ellipses (purple, red, orange, respectively) of the best-fit parameters in the Burkert halo case. The central points indicate the best-fitting values. $\frac{M}{L}$ is in the  IRAC 3.6 $\mu$m band units.}

\label{ELLBUR}

\end{centering}
\end{figure*}

\noindent
Then, we compute the mass of the stellar disk as
\begin{eqnarray}
M_\mathrm{D} \simeq 1.1 {\Upsilon_*^{3.6}(V_{\mathrm{d}})(R_{\mathrm{l}})^2} \frac{R_{\mathrm{l}}}{G}
\label{md}
\end{eqnarray} 
 
\noindent
 where $V_{\mathrm{d}}(R_{\mathrm{l}})$ is the disk contribution to the circular velocity at the outermost radius $R_{\mathrm{l}} \approx 48$ kpc. This estimate is very solid and independent of the actual light profile in the inner part of the galaxy. We find $M_\mathrm{D} \simeq 4.4 \times 10^{10} M_{\odot}$ \citep[$\approx 2$ times bigger than the values found by][]{bloketal} with a propagated uncertainty of about ten percent. The corresponding virial mass and virial radius are $M_\mathrm{vir}=5.8^{+0.4}_{-0.8} \times 10^{11} M_{\odot}$ and $R_\mathrm{vir}=214^{+4}_{-11}$ kpc.The 1,2,3-$\sigma$ confidence regions for the best-fit parameters are shown in Fig. \ref{ELLBUR}. The central points correspond to the best-fit values.
 
\begin{figure*}
\begin{centering}
\makebox[\textwidth][c]{

\includegraphics[width=0.225\textwidth, angle=0]{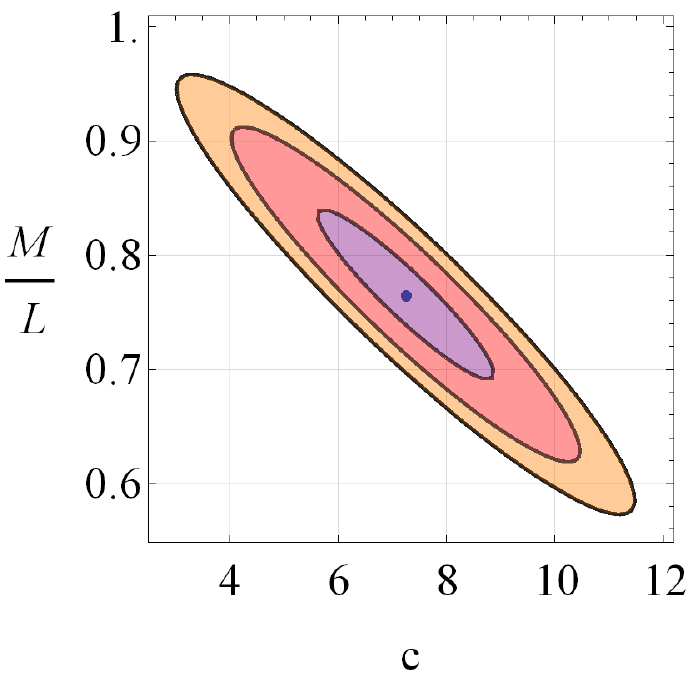}
\includegraphics[width=0.218\textwidth, angle=0]{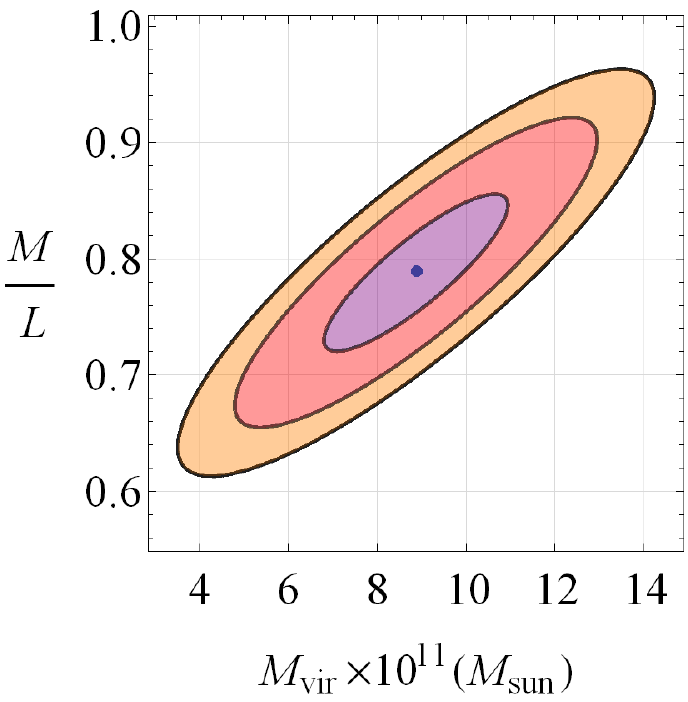}
\includegraphics[width=0.2\textwidth, angle=0]{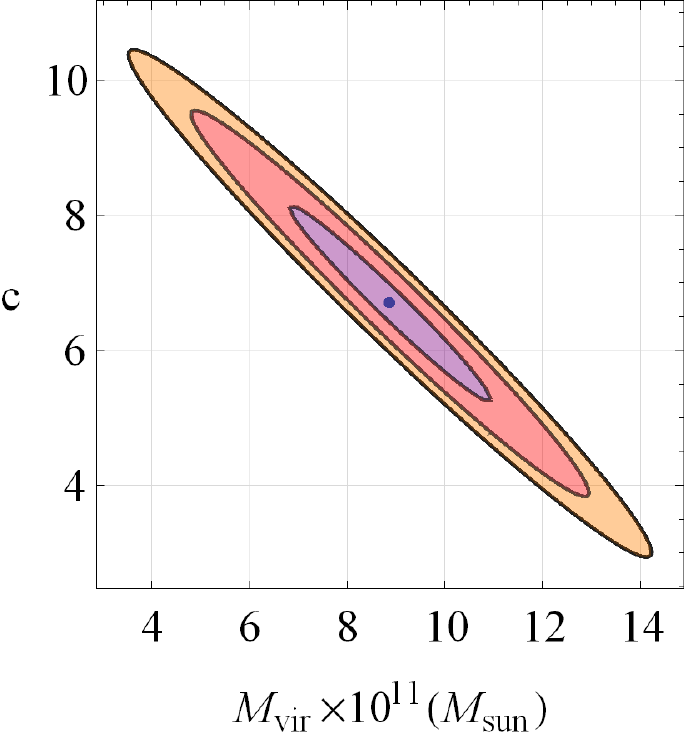}

}
\caption{ 1, 2, 3$\sigma$ confidence ellipses (purple, red, orange, respectively) of the best-fit parameters in the NFW halo case. The central points indicate the best-fitting values. $\frac{M}{L}$ is in the  IRAC 3.6 $\mu$m band units.}

\label{ELLNFW}
\end{centering}
\end{figure*}

In the framework of  the NFW mass models, we fitted data in terms of the free parameters: the virial mass, the concentration parameter and  above defined the mass-to-light ratio ($M_\mathrm{vir}, c, \Upsilon_*^{3.6}$). The results of the best-fit are
\newline
\indent
\begin{eqnarray}
M_\mathrm{vir} &=& (8.9 \pm 2.1) \times 10^{11} \quad \mathrm{M_{\odot}};\nonumber\\
c &=& (6.69 \pm 1.46);\nonumber\\
\Upsilon_*^{3.6} &=& (0.79 \pm 0.07).\nonumber
\end{eqnarray} 

\noindent
In this case the reduced chi-square is ${\chi}^2=0.8$, even slightly better value than found for  the URC-halo model.
The best-fit value of the concentration parameter $c= 6.69\pm 1.46$ is found to be somewhat lower than what is expected from Eq. (\ref{cdm}): $c_{\mathrm{NFW}} \approx 10 \pm 1$.  It is worth recalling that in other galaxies, this discrepancy in the concentration parameter is much larger \citep[see][]{McGaugh, salucci10, memola}. 
  
From Eq. (\ref{md}) the disk mass, within a ten percent uncertainty, is $M_{\mathrm{D}}=3.5 \times 10^{10} M_\odot$, a somewhat smaller value than found for the URC-halo model. The best-fit values of the scale radius and the characteristic density are $r_\mathrm{s} = (37.2 \pm 11.0)$ kpc and $\rho_\mathrm{s} = (8.0 \pm 4.1) \times 10^{-26}$ $\mathrm {g\ cm^{-3}}$.
\newline
\indent
The 1,2,3-$\sigma$  best-fit parameters confidence regions are shown in Fig. \ref{ELLNFW}. The central points correspond to the best-fit values that result somewhat higher or lower than the N-Body simulation outcome relative to a galaxy with $V_{max}\simeq 150 \ \mathrm{km/s} $ as NGC 3198. The discrepancy, however, is within 1.5 $\sigma$.\

The standard mass modelling of the kinematical data of NGC 3198 is then not able to clearly select between a cored and a cuspy halo profile. In fact, in the case where a galaxy like NGC 3198 showing  a flattish rotation curve over a wide range of radii, we have a modelling degeneracy:  the same best-fit solution corresponds to very different mass models  \citep[see Appendix of][]{gentile04}.

\section{A NEW METHOD OF ESTIMATING THE HALO DM DENSITY AND ITS RESULTS}

A step forward in mass  modelling  spirals has come from the method of   \citet{salucci10}. This method was applied first to the Milky Way  to derive the value of  the DM density at the Sun's location. In this paper we are applying it, for the first time, to the outermost parts of an external galaxy. Our aim is to derive, for the spiral  galaxy with the most extended kinematics, the DM density at large radii, where the influence of the stellar  and  HI disks is respectively negligible and  known. 

The idea underlying the \citet{salucci10} method is to resort to the equation of centrifugal equilibrium holding in spiral galaxies:

\begin{equation}
\frac{V^2}{r} = a_\mathrm{H}+a_\mathrm{D}+a_\mathrm{HI}
\label{accel}
\end{equation}

\noindent
where $a_\mathrm H$, $a_\mathrm D$ and $a_\mathrm{HI}$ are the radial accelerations, generated, respectively, by the halo, stellar disk, and HI disk mass distributions. Within the approximation of spherical DM halo, we have 

\begin{equation}
a_\mathrm{H} = 4 \pi G r^{-2} \int\limits_0^r \rho_{H}(r) r^{2}\,dr.
\label{accelh}
\end{equation}

\noindent
Therefore, we compute the derivative of Eq. (\ref{accel}) by manipulating previous equations to get the DM density at any radius. We have
\\
\begin{equation}
\rho_\mathrm{H}(r) = \frac{X_\mathrm q}{4 \pi G r^{2}} \frac{d}{dr} \Biggl\lbrack r^{2} \Bigl( \frac{V^{2}(r)}{r}-a_\mathrm{D}(r)-\frac{V_\mathrm{HI}^{2}}{r} \Bigr) \Biggr\rbrack
\label{density}
\end{equation}
\\
\noindent
where $X_\mathrm q$ is a factor correcting the spherical Gauss law used above in case of any oblateness of the DM halo. Since this value is very near to one, we assume $X_\mathrm{q}=1$  \citep[see details in][]{salucci10}.

  Equation(\ref{density}) gives a very good estimation of the density when the contribution from the luminous components is small.  In short, to work well the new method  requires a high-resolution, high-quality, and very extended kinematics for a spiral of known distance. Of course, we are  interested in the region well outside $3 R_D$.  

For simplicity, we model the disk component as a Freeman stellar exponential, infinitesimally thin disk \citep{freeman} with the disk  scale length $R_{\mathrm{D}}=3.7$ kpc. No result in this paper will change by instead assuming the  \citet{bloketal} disk mass profile, whose contribution to the  circular velocity $V_D$  is given in Table \ref{tbl:1}. For $M_\mathrm D$ we use the URC mass modelling value: $M_\mathrm D \simeq 4.4 \times 10^{10} M_{\odot}$, and no result changes if assuming any other reasonable value for this quantity (see below). 

The surface stellar density profile is given by
  
\begin{equation}
\Sigma(r) =\left (\frac{M_\mathrm{D}}{2 \pi R_\mathrm{D}^2}\right) e^{-\frac{r}{R_\mathrm D}}.
\label{sigma}
\end{equation}

\begin{figure}
\begin{centering}
\includegraphics[angle=0,height=5truecm,width=9truecm]{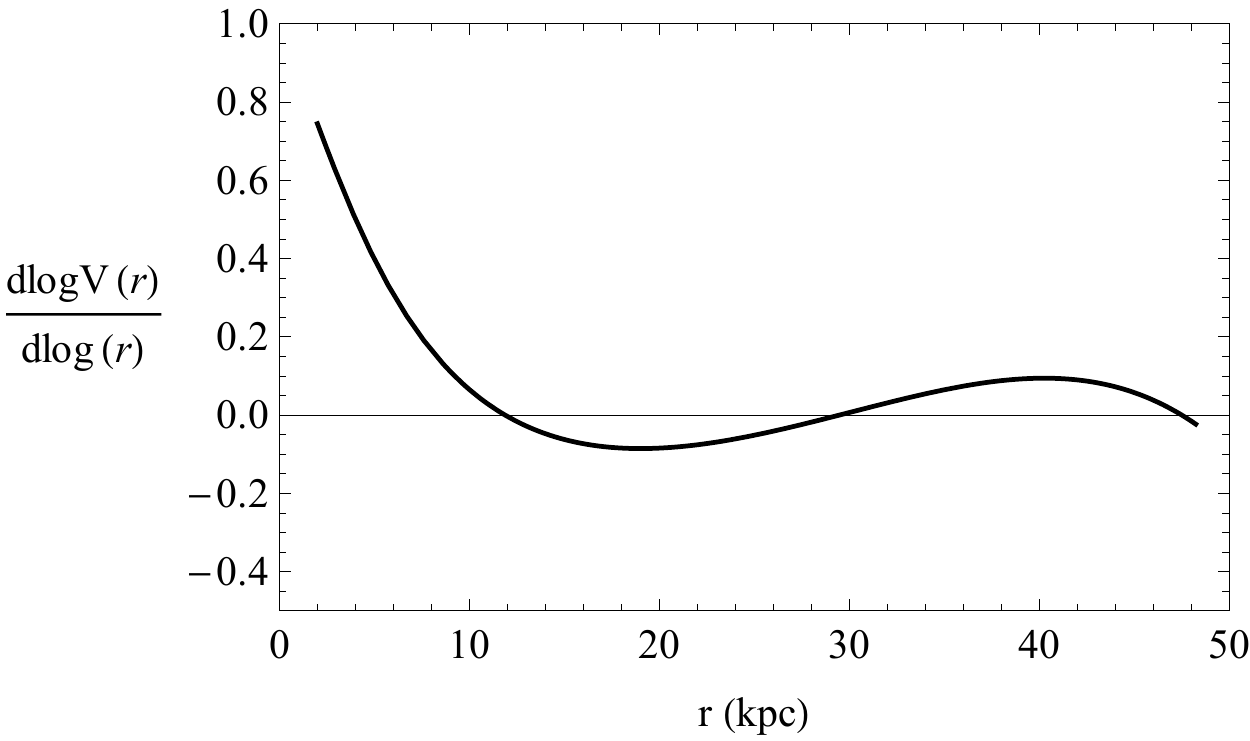}
\caption{Logarithmic slope of the RC of NGC 3198.}
\label{slope}	
\end{centering}
\end{figure} 

\noindent
 
 We can write
 \\
\begin{equation}
a_\mathrm{D}(r) = \frac{G M_\mathrm{D} r}{R_\mathrm{D}^3}(I_{0}K_{0}-I_{1}K_{1})
\label{fremandisk}
\end{equation}
\\
\noindent
where $I_\mathrm n$ and $K_\mathrm n$ are the modified Bessel functions computed at $r/(2 R_ 
\mathrm D)$.

The  HI disk component $V_{\mathrm{g}}(r)$ and its derivative  are easily obtained from observations \citep{gentileetal}.
Finally, Eq. (\ref{density}) becomes

\begin{eqnarray}
\rho_\mathrm{H}(r) &=& \frac{1}{4 \pi G} \Bigl\lbrack \frac{V^{2}(r)}{r^2}(1+2 \alpha(r))-\frac{G M_\mathrm D}{R_\mathrm{D}^3} H\left(\frac{r}{R_\mathrm D}\right)\nonumber \\
&-& \frac{V_\mathrm{HI}^{2}(r)}{r^2}(1+2 \gamma(r)) \Bigr\rbrack
\label{finden}
\end{eqnarray}
\\

\noindent
where $2 H\left(\frac{r}{R_\mathrm D}\right) = (3 I_0 K_0 - I_1 K_1) + \frac{r}{R_\mathrm D}(I_1 K_0 - I_0 K_1)$ and $\alpha(r)$ and $\gamma(r)$ are the logarithmic slopes of the circular velocity and of the HI+He disk contribution to the latter, both of which known. 
\newline
 \begin{figure*}
\begin{centering}
\includegraphics[angle=0,height=8truecm,width=13truecm]{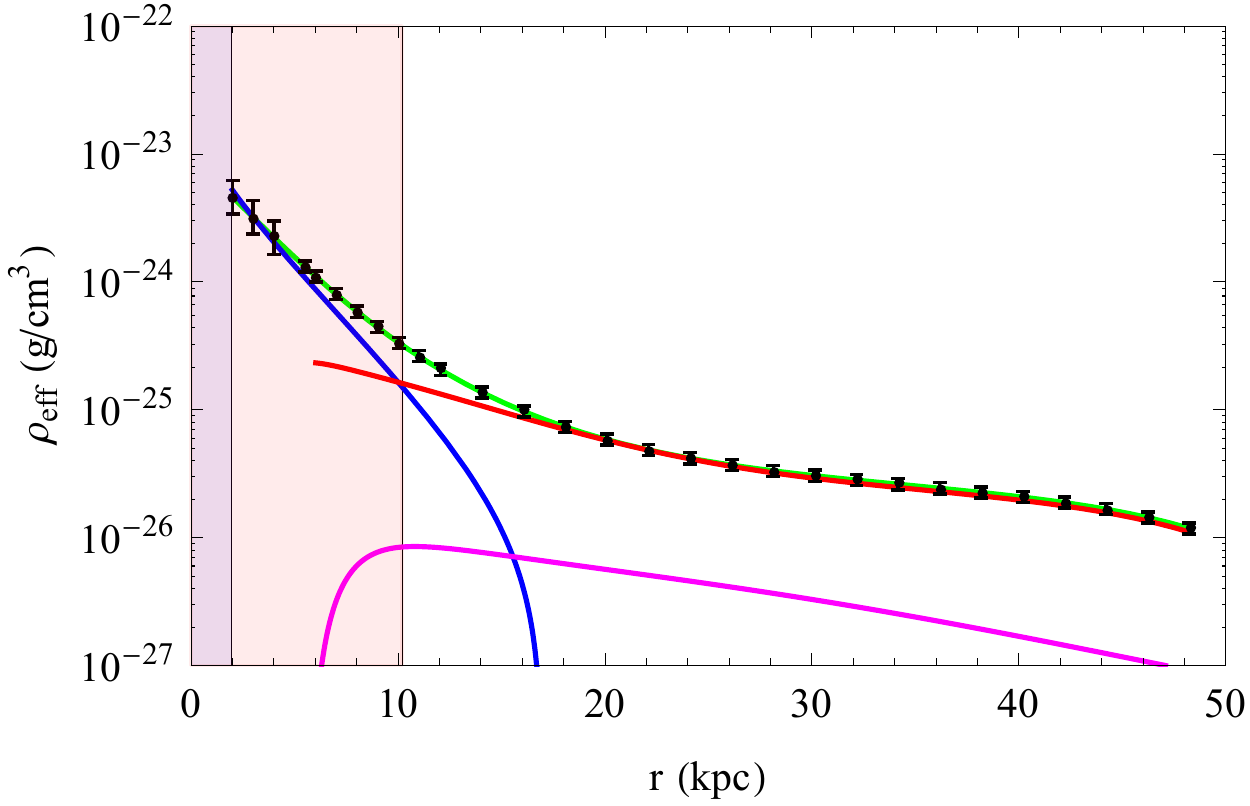}
\caption{Density profile of the DM halo of NGC 3198 and the \emph{effective} density of the other components. We assume $M_\mathrm{D}=4.4 \times 10^{10} M_{\odot}$. The stellar disk (Blue line), the HI disk (magenta line), the dark halo (red line) and the sum of all components (green line). The different colour regions correspond to the regions a) where we do not have any kinematical information due to the lack of data (dark purple), b) where the stellar disk dominates the DM density profile (light purple) and c) where DM dominates (white).}
\label{densities}	 
\end{centering}
\end{figure*} 
\newline
\begin{figure*}
\begin{centering}
\includegraphics[angle=0,height=8truecm,width=11truecm]{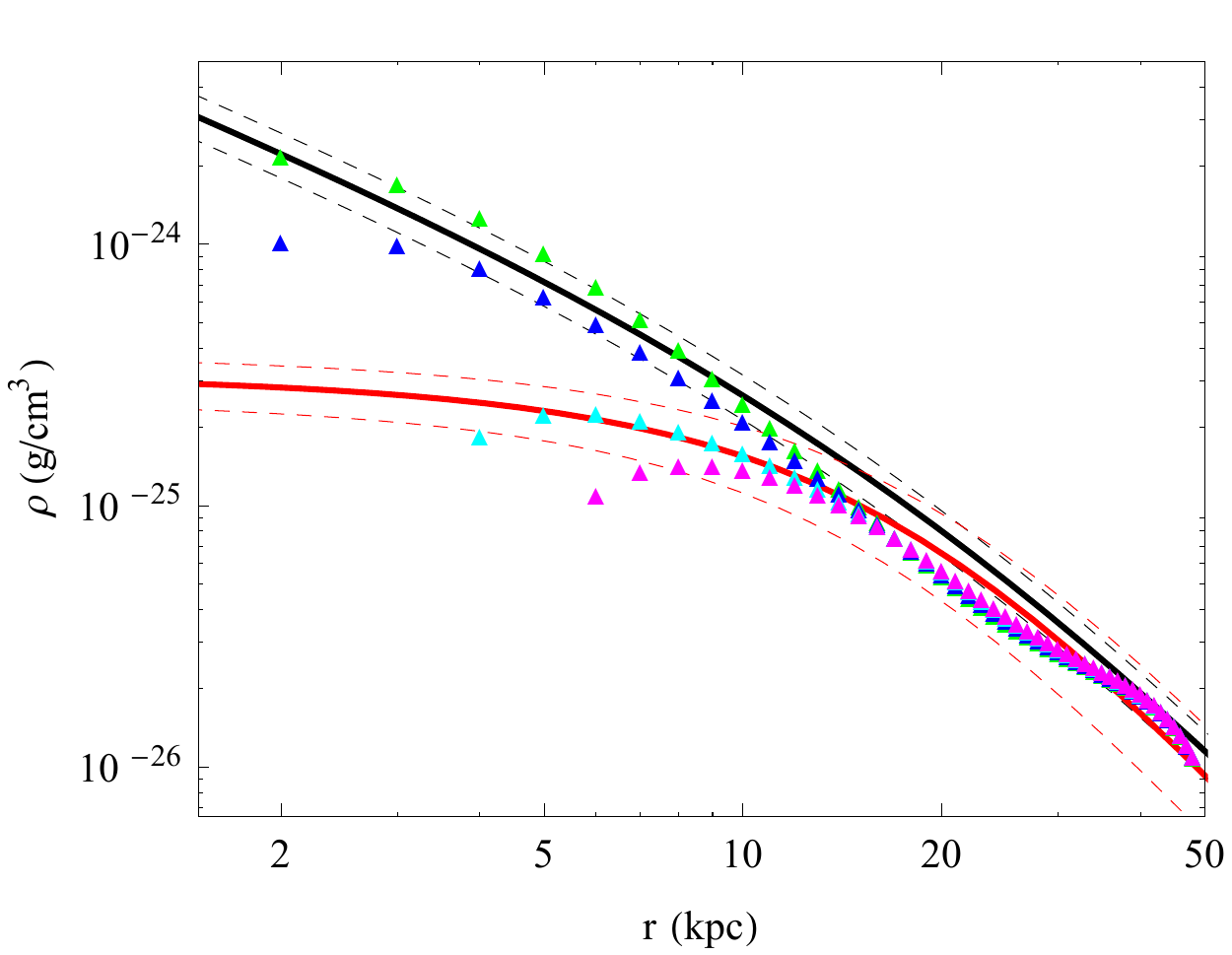}
\caption{DM density profiles for different disk mass values $M_\mathrm{D}=n \times 10^{10} M_{\odot}$ and n=2,3,4.4,5  from the highest to the lowest curve (green, blue, azure and magenta triangles). The black solid line shows the NFW density profile with $M_\mathrm{vir}=8.9 \times 10^{11} M_{\odot}$ corresponding to the best fit values found in the previous section. Two black dashed lines show the NFW density profiles taking $1\sigma$ uncertainties in $M_\mathrm{vir}$ and 10 percent of uncertainties in $c$ into account. The URC halo is shown as Red solid line. Two Red dashed lines show the URC density profiles obtained by taking $1\sigma$ uncertainties in $\rho_\mathrm{0}, r_\mathrm{c}$ into account.}
\label{profiles}	
\end{centering}
\end{figure*} 
\newline
 \begin{figure*}
\begin{centering}
\includegraphics[angle=0,height=8truecm,width=12truecm]{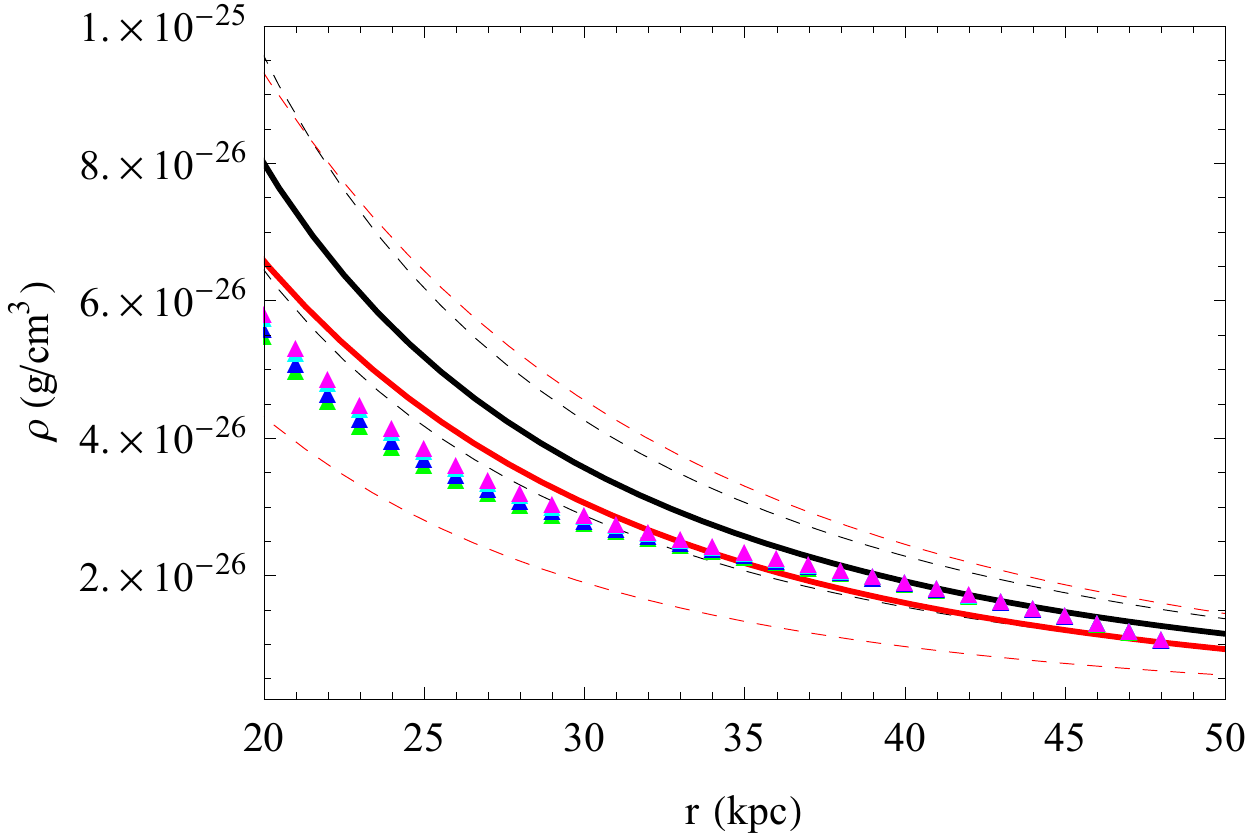}
\caption{Zoom of Fig. \ref{profiles} on linear scales.}
\label{zoomprofiles}	
\end{centering}
\end{figure*} 
\newline
\indent
We stress that in galaxies with  $V_d(3 R_\mathrm{D}) \simeq V(3R_\mathrm{D})$, it is very difficult to use the standard mass modelling method to disentangle  the  circular velocity into its dark and luminous  components and to obtain the DM density distribution out to $r\simeq6 R_\mathrm{D}$.  Instead,  the fundamental point of the present new method  is that,  for radii $r \gtrsim 3 R_\mathrm{D}$ the second term of RHS of Eq. (\ref{finden}) always  goes rapidly to zero becoming much smaller than the first  and the third terms, both known. Then, by means of Eq.  (\ref{finden}), we can immediately derive $\rho(r)$: the unknown term, proportional to the stellar disk mass becomes irrelevant as $r\gtrsim3 R_\mathrm{D}$. In Eq. (\ref{finden}), all  terms have the dimensions of a density; specifically, the three terms of  RHS can be considered as the \emph{effective} density of the whole gravitating matter and  the (sphericized) densities  of the stellar and gaseous disks.  

Thus,  by means of  Eq. (\ref{finden})  for   $r\gtrsim 3 R_\mathrm{D}$, we obtain  a reliable  dark matter density profile, the farther we get, the more precise the estimate becomes. In the range : $r \lesssim 3 R_\mathrm{D}$, Eq. (\ref{finden}) still holds,  but  it cannot give information about the DM halo since this is very sub-dominant there. In this region, however,  we can use the latter equation to derive the stellar disk mass. This estimate, however, may turn out to be somewhat uncertain  because, in this inner region, the first term of the RHS of Eq. (\ref{finden}) has some observational errors; moreover,  moderate errors  in the measured value of the disk scale length can affect the third term. Incidentally, we notice that the distance of the galaxy must be  known with good precision (as it is in NGC3198), because its uncertainty affects all three terms of the RHS in different ways.

 In the case of NGC 3198 we take $\gamma(r)=0$ for  $r\gtrsim3 R_\mathrm{D}$, in order to simplify our calculations, since the gas contribution to the circular velocity, from $\thicksim 12$ kpc, is nearly constant (see Fig. \ref{BURKERT}). The logarithmic slope of the RC instead varies with radius  (see Table \ref{tbl:1}), even at   $r \thicksim 17$ kpc: we take $\mathrm{d}\log \mathrm{V(r)}/\mathrm{d}\log \mathrm{r}$ as in Fig. \ref{slope}. In Fig. \ref{densities} we obtain the density profile of the dark halo. We list the values of the obtained DM profile starting from 5.5 kpc in Table \ref{tbl:1}. We see  that the DM component starts to dominate the luminous components from  $r \thicksim 10$ kpc; moreover, the stellar disk's contribution in the  RHS of Eq.  (\ref{finden}) goes further below the gravitating matter for $r \gtrsim 10$ kpc and to zero for $r \gtrsim 17$ kpc, independently of its mass. This means that, starting from $\thicksim 17$ kpc, the halo density profile, which obtained by means of  Eq. (\ref{finden}), is virtually  free from the uncertainty on the actual value of the disk mass,  which usually plagues the standard mass modelling of  RCs. 

Next we discuss in detail the distribution of matter in the various regions of NGC 3198. In the innermost one, $r \lesssim 2$ kpc we do  not have any kinematical information  due to the lack of data. In  the region extended  from $\thicksim 2$ kpc to $\thicksim 10$ kpc,  the stellar component dominates over the DM component.  No direct information on the latter can be extracted here. In this region, however,  we can use the fair  agreement between the stellar  and the dynamical density (as defined in  Eq. (\ref{finden})) to derive the value of $M_D$.  
 
For  $r \gtrsim 17$ kpc, the DM density is directly obtained by means  of Eq. (\ref{finden}) see Fig. \ref{densities}. Here, all quantities entering in Eq. (\ref{finden}) are known within a small uncertainty,  and this  leads to a robust determination of $\rho(r)$. 
  
The effect of the uncertainty in the value of the disk mass in the determination of $\rho(r)$  is shown  well in Fig. \ref{profiles}. We now do not  take a specific value for $M_D$,  and  we derive the DM density profiles  through Eq. (\ref{finden})  by assuming different values for the disk mass.  We see that the \citet{salucci10} method leads to a family of density profiles that all agree outside the stellar disk (i.e. for $r\gtrsim 3 R_\mathrm{D}$), independently of the corresponding value of the disk mass (see the zoomed area in Fig. \ref{zoomprofiles}) .

\begin{figure*}
\begin{centering}
\includegraphics[angle=0,height=8truecm,width=8truecm]{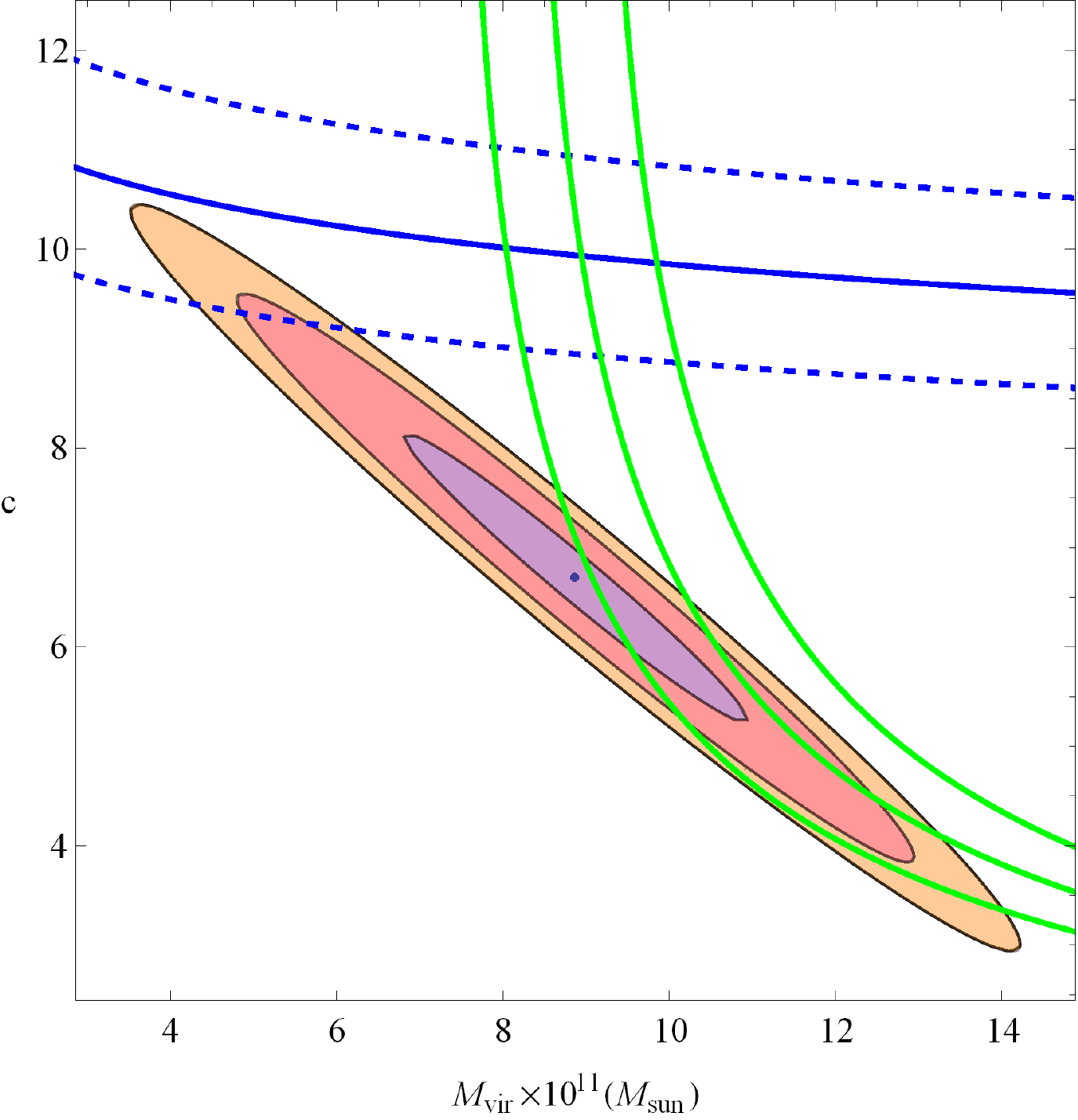}
\caption{NFW case: the 1,2,3 - $\sigma$ (purple region, red region and orange region, respectively) confidence ellipses for the global best-fit parameters. The blue solid line shows the $c-M_\mathrm{vir}$ relation from numerical simulations, the dashed blue lines show its 10 percent uncertainty. The Green region shows the $c-M_\mathrm{vir}$ relation from the local density values obtained by the \citet{salucci10} method taking the 10 percent uncertainty into account.}
\label{sim}	
\end{centering}
\end{figure*} 

We compare the derived  DM density  with the URC and the NFW profiles, see Fig. \ref{profiles}. In the first case, the derived  density  bears  no difference with that obtained by means of  the first method, i.e. with Eqs.(\ref{vel}) and (\ref{urc}). We therefore found $M_D=(4.4\pm1.0)\times10^{11}M_{\odot}$.  This is also evident in the zoomed area of Fig. \ref{profiles} (see Fig. \ref{zoomprofiles}). In  external regions of  the NGC 3198, the \citet{salucci10}  method  yields a halo that is compatible with the URC halo, as derived by chi-square fitting of the RC of  NGC 3198.

For NFW haloes the situation is very different. First, in Fig. \ref{zoomprofiles} we realize that,  independently of the disk mass we assume, the best-fit NFW halo profile  is in poor agreement with the derived density. We consider this fault only as a hint and not as a definitive evidence against a NFW halo. Within the new modelling method, we derive  the value of $c$ and $M_\mathrm{vir}$ and their uncertainties by evaluating,  from  Eqs.(\ref{nfw}), (\ref{rho}), and (\ref{finden}), the DM density  (and its uncertainties) at the radius $r=45$ kpc (where the stellar disk certainly  does not contribute to the gravitating matter density profile). A more serious problem appears when we realize that the resulting values of the halo parameters $M_\mathrm{vir}$ and $c$  are very different from the ones derived by means of  the standard method  applied in the previous section.  In fact, for  the virial mass and  the concentration parameters we get $M_\mathrm{vir}=(8.9 \pm 2.1) \times 10^{11}  M_{\odot}$, $c=6.69 \pm 1.46$. 
\newline
\indent
In Fig. \ref{sim} we plot the  two solutions for $(c-M_\mathrm{vir})$  with their uncertainties. We now  compare these values of the concentration parameter and the virial mass $(c-M_\mathrm{vir})$ obtained by each of the mass modelling techniques also in light of the  $(c-M_\mathrm{vir})$ relationship that emerged from numerical simulation (see Eq. (\ref{cdm})).
 They  agree only for  values of $c$ that are much lower than those  emerging in numerical simulations and for values of  $M_{\mathrm{vir}}>9\times 10^{11} M_{\odot} $   far too high for this spiral galaxy  which has  $V_\mathrm{max}\lesssim150 $ km/s.
 \newline
 \indent
In short, by assuming a NFW halo in NGC 3198, we have two completely {\it different} solutions for its structure parameters, according to whether one adopts the local or the global method of mass modelling.  The best intersection of these sets of results is  in total disagreement with the properties of the simulated haloes in the N-body $\Lambda$CDM scenario. Noticeably, the problem for NFW haloes is different here and, if possible,  more serious  than that of the  core-cusp discrepancy, usually occurring at $\sim 0.05 R_\mathrm{vir}$ \citep{donato}. We found, in fact, that the density of the DM halo around NGC 3198 is not very consistent with the NFW profiles well out to $0.22 R_\mathrm{vir}$. 

\begin{figure*}
\begin{centering}
\includegraphics[angle=0,height=8truecm,width=11truecm]{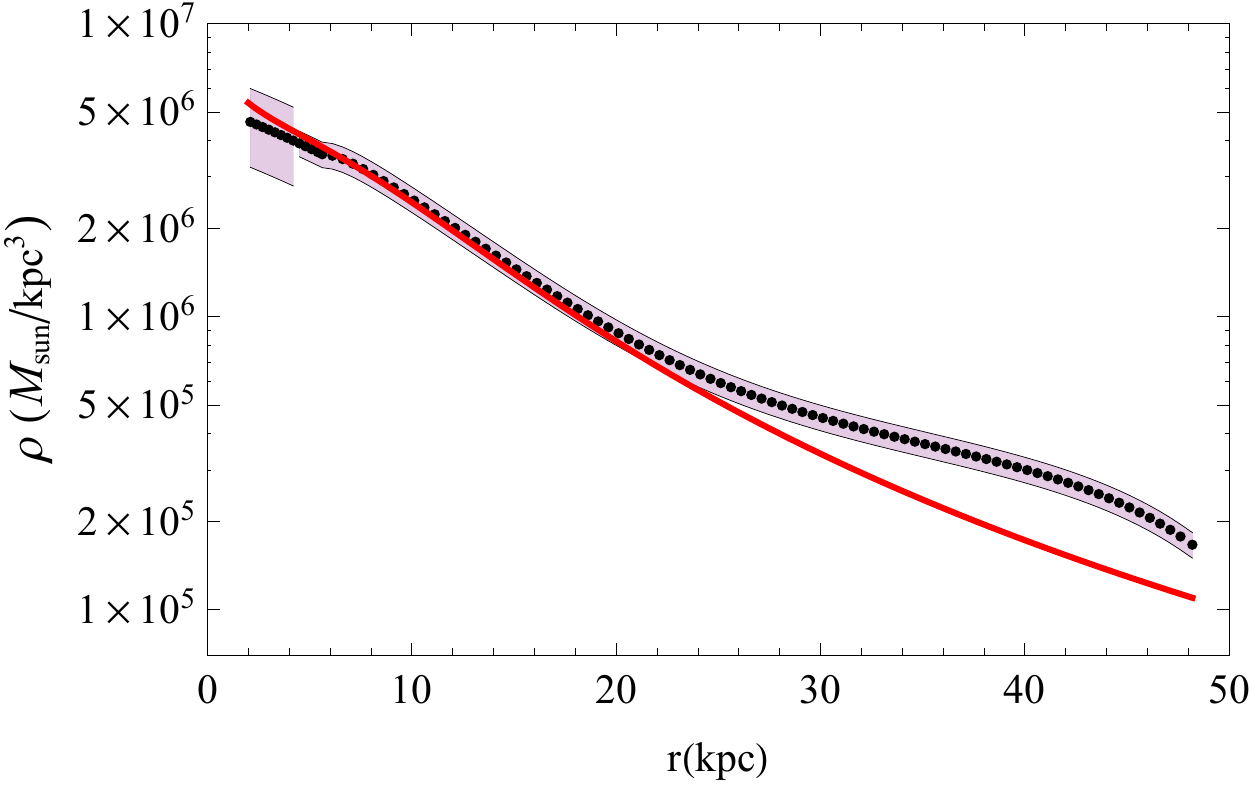}
\caption{Observed  DM halo density profile (black points) of NGC 3198 and the obtained errors by the error propagation analysis (purple area). \citet{dicintio} mass-dependent model predictions (red line).}
\label{fitden}	 
\end{centering}
\end{figure*} 

 Of course, NFW haloes  emerge out of  DM only simulations in the leading   $\Lambda$CDM scenario. However, this scenario actual haloes around galaxies may have undergone  a compression by the stellar disk during the   formation of the latter \citep[e.g.][]{gnedin,katz} and/or suffered  a baryonic  feedback during the subsequent history of the galaxy \citep{dicintio}. These processes could have  modified  the DM haloes original distribution.  Furthermore,  haloes today around galaxies, can be born  within a different cosmological  scenario like  Warm or Self-Interacting DM,  in which a NFW profile  is not  established \cite[e.g.][]{devega}.

In spite of that,   in this section  we  focused on the failure of the NFW halo profile because:  a) it is important  to  fully expose the discrepancy between the density distribution of the dark haloes around galaxies and the predictions of the simplest  dark particle scenario;  b)  NFW haloes are still used often to investigate  important Cosmological issues in the  belief that  the discrepancy with actual galaxy haloes, though present, does not have much physical relevance; c)  de facto, several cosmological  investigations have been carried out considering that  there is no discrepancy at all. 
\newline
\indent
 It goes without saying that the failure of NFW profiles is not the demise of the $\Lambda$CDM scenario (see next section).

\section{Testing LCDM  halo profiles modified by the physics of stellar disk formation}     
                                                       
 A  line of thought holds that  the cosmological  core-cusp problem (CCCP)  can be addressed by considering the  present DM haloes around galaxies like NGC 3198 as very different with respect to those emerging out of N-body simulations. In fact, the formation and the growth in them of stellar disks  and the related numerous and powerful  supernova explosions could have modified the original N-body profile,  by making it shallower and more in agreement with observations. We stress that  such a dissolution of the CCCP is rarely studied.  

One exception is the work by \citet{dicintio} based on the analysis of hydro-dynamically simulated galaxies drawn from the MaGICC project \citep{brook,stinson}. They did find, at the end of these simulations, that DM haloes had a completely new family of profiles, a generic double power-law density profile \citep{dicintio}. This was found to vary in a systematic manner in the stellar-to-halo mass ratio of each galaxy.
 
The mass-dependent density profile was derived by analysing  hydrodynamical cosmological simulations.  This profile (hereinafter referred to as DC14) accounts for the effects of feedback on the DM haloes due to gas outflows generated in high density star-forming regions during the history of the stellar disk. The resulting radial profile is far from simple, since it starts from an ($\alpha, \beta,\gamma$) double power-law model \citep[see][]{dicintio}  

\begin{eqnarray}
\rho_\mathrm{DC14}(r)=\frac{\rho_s}{(\frac{r}{r_s})^{\gamma}\left(1+(\frac{r}{r_s})^{\alpha}\right)^\frac{(\beta-\gamma)}{\alpha}}
\label{DC14}
\end{eqnarray}
\\
where $\rho_s$ is the scale density and $r_s$ the scale radius. The inner and the outer regions have logarithmic slopes $-\gamma$ and $-\beta$, respectively, and $\alpha$ indicates the sharpness of the transition. These three parameters are fully constrained in terms of the stellar-to-halo mass ratio as shown in \citet{dicintio}:

\begin{eqnarray}
\alpha&=&2.94-\mathrm{log}_{10}\lbrack(10^{X+2.33})^{-1.08}+(10^{X+2.33})^{2.29}\rbrack\nonumber\\
\beta&=&4.23+1.34X+0.26X^2\\
\gamma&=&-0.06+\mathrm{log}_{10}\lbrack(10^{X+2.56})^{-0.68}+(10^{X+2.56})\rbrack\nonumber
\label{slopes}
\end{eqnarray}
 \\
 where $X=\mathrm{log}_{10}\left(\frac{M_{star}}{M_{halo}}\right)$.
                                                                          
The concentration parameter of the halo is $c=\frac{R_\mathrm{vir}}{r_s}$. An alternative definition, adopting the radius $r_{-2}$, is $c_\mathrm{DC14}=\frac{R_\mathrm{vir}}{r_{-2}}$, where $r_{-2}$ is the radius at which the logarithmic density slope of the profile is -2. This definition allows defining the same physical $r_s$ for different values of  ($\alpha, \beta,\gamma$) 

\begin{eqnarray}
r_{-2}=\left(\frac{2-\gamma}{\beta-2}\right)^{\frac{1}{\alpha}}r_s.
\label{radius}
\end{eqnarray}

Following \citet{dicintio}, we reach the relation between the concentration parameter in the hydrodynamical simulations and the N-body  haloes  as a function of the stellar-to-halo mass ratio:

\begin{eqnarray}
c_\mathrm{DC14}=(1.0+0.00003 e^{3.4(X+4.5)})\times c_\mathrm{NFW}.
\label{concent}
\end{eqnarray}
\\

Using the definition of the enclosed mass, we can write down the expression for the scale density of the DC14 profile:

\begin{eqnarray}
\rho_{s}=\frac{M_\mathrm{vir}}{4 \pi \int\limits_0^{R_\mathrm{vir}}\frac{r^2}{(\frac{r}{r_s})^\gamma\lbrack1+(\frac{r}{r_s})^\alpha\rbrack^{\frac{\beta-\gamma}{\alpha}}}\mathrm{dr}}.
\label{scaleden}
\end{eqnarray}

To reduce the number of free parameters of the DC14 profile, it is necessary to adopt a value for the concentration parameter of the original NFW halo. For our purpose and since this quantity depends extremely weakly on the virial mass,  we can adopt:  $c_\mathrm{NFW}=10$. By bringing together all the above equations  of this section and noticing that  $R_\mathrm{vir}=c_\mathrm{DC14}r_{-2}$, we can rewrite Eq. (\ref{DC14}) just as a function of the scale radius and the stellar-to-halo mass ratio.

With this density profile, we can attempt a two-free parameter ($X,r_s$)  fit of  the {\it derived} halo density profile of the  NGC 3198  (Fig. \ref{densities}). The result is in Fig. \ref{fitden}, the best-fit parameters are:

\begin{eqnarray}
X&=&-2.6;\nonumber\\
r_s&=&12.2 \mathrm{kpc}.\nonumber
\end{eqnarray}

The inferred virial radius is $R_\mathrm{vir} \simeq 235 \mathrm{ kpc}$ and virial mass is $M_\mathrm {vir} \simeq 7.7 \times 10^{11} \mathrm {M_{\odot}}$.
As result, the CCCP is only partially resolved, and the DM core we detect in NGC 3198 can be explained  as an effect  of the energy  that, in different ways,  stars have injected in the galactic  ambient. However,  in the outer densities of the dark halo derived for the first time in this study, there are features that may conflict with the N-body simulation predictions that should be  recovered at $R>25$ kpc from the centre of NGC 3198.

\section{Conclusions}

Galaxies with a flattish rotation curve between 5 kpc to 50 kpc (e.g. the spiral  NGC 3198, for many years the flagship of  the evidence of the DM in galaxies) amount to only a few percent of the total number of disk systems.  However, they  play a crucial role in the core-cusp issue of the DM density.  In fact,  these  "flat" RCs  can be  well  fitted by either a cored DM halo or a cuspy DM halo, just by adjusting the amount of the stellar matter content. In contrast, in spirals in which  $|\mathrm{d}\log \mathrm{V(r)}/\mathrm{d}\log \mathrm{r}|$  is significantly far from zero, this quantity  constraints  the distribution of DM, usually towards a cored one \citep{gentile05}. 

NGC 3198 is a special galaxy. The HI disk of this galaxy is very extended out to $\sim$ 13 disk  scale lengths or out to $\sim$ 0.22 $R_\mathrm{vir}$. Furthermore, for the innermost region, many complementary optical kinematical measurements are available.  This object (of known distance  of 13.8 Mpc) shows a spectacular evidence of a dark force in action: baryons of this galaxy  are clearly unable to account for its (very extended)  kinematics. We must assume that a large part of the circular velocity of NGC 3198  is due to a DM halo. In addition,  the circular velocity of this galaxy is at variance with the MOND paradigm \citet{ gentileetal}, while  it seems  plausible within the F(R) scenario \citep[see][]{fr}. It is then obvious that to resolve the core-cusp issue in NGC 3198 is of particular importance.

To do so, we used the old optical and new HI kinematics performed in the HALOGAS survey in combination with a new method of mass modelling a rotation curve. First, we verified that with the standard  ${\chi}^2$ mass modelling of the RC,  both  URC Burkert and the  NFW dark halo models  fit the available data well. Then, once assured  of the conditions of its applicability,  we used the new refined method developed by \citet{salucci10} to determine  the  DM halo density of  NGC 3198.  This result is inconsistent  with  the NFW haloes predictions, independently of any assumption about the luminous component we can take.  Moreover,  the derived density profile strongly supports  a cored distribution of DM, adding independent evidence to the idea of cored DM distribution in galaxies.

 Today, within the $\Lambda$CDM scenario,  NFW haloes are often still assumed, although it has been recognized  that in this scenario the actual DM halo density is not the one emerging from cosmological N-body simulations. In fact, it has been agreed that they have been modified by the subsequent baryonic physics, leading to stellar disk formation and evolution. \citet{dicintio} have simulated this important phase of the history of spirals and found the emerging profiles of dark matter haloes. This baryonic $\Lambda$CDM halo profile prediction fits the detected halo of NGC 3198 very well, especially in its cored region. At very large distances, 25 kpc, however,  the DM halo density derived here results in a clash; i.e., it  is significantly {\it higher} than the outcome of the hydrodynamic N-body $\Lambda$CDM simulations. This disagreement is not an isolated one \citep[see][]{gentile07}.

\begin{acknowledgements} 
We would like to thank the referee, Albert Bosma, for comments
that improved the quality of the paper.
 \end{acknowledgements}


\bibliographystyle{aa} 
\bibliography{N3198A&Areferee} 

\end{document}